\documentclass[a4paper,11pt]{article}

\pdfoutput=1
\usepackage{fullpage}

\usepackage{jheppub}
\usepackage{color}
\usepackage{graphicx}
\usepackage{tabularx}
\usepackage{xspace}
\usepackage{subcaption}
 \usepackage{tikz}
\usepackage[compat=1.1.0]{tikz-feynman}

\usepackage{float}

\usepackage{verbatim}
\usepackage{amsmath}
\usepackage{amssymb}
\usepackage{url}
\usepackage{bbold}
\usepackage{slashed}
\usepackage{array}

\usepackage{multirow}
\usepackage{threeparttable}
\usepackage{paralist}
\usepackage{xspace}
\usepackage{upgreek}
\usepackage{lipsum}

\newcommand{\dbar}[1]{\dfrac{\mathrm d #1}{(2\pi)}}
    \newcommand{\be}{\begin{equation}}
  \newcommand{\ee}{\end{equation}}
    \newcommand{\ba}{\begin{align}}
  \newcommand{\ea}{\end{align}}

\newcommand{\du}{{\rm d}}

\newcommand{\kpc}{\textrm{kpc}}
\newcommand{\pc}{\textrm{pc}}

\newcommand{\GeV}{\textrm{GeV}}
\newcommand{\MeV}{\textrm{MeV}}
\newcommand{\cm}{\textrm{cm}}
\newcommand{\pb}{\textrm{pb}}
\newcommand{\km}{\textrm{km}}
\newcommand{\mtr}{\textrm{m}}
\newcommand{\s}{\textrm{s}}
\newcommand{\sr}{\textrm{sr}}
\newcommand{\Myr}{\textrm{Myr}}

\renewcommand{\L}{\mathcal{L}}
\newcommand{\nbar}{\bar n}
\renewcommand{\dbar}{\bar d}
\newcommand{\pbar}{\bar p}
\newcommand{\AMS}{\textit{AMS-02}\;}
\newcommand{\GAPS}{\textit{GAPS}\;}

\title{Searching for Dark Photon Dark Matter with Cosmic Ray Antideuterons}
\author{Lisa Randall, Weishuang Linda Xu}

\affiliation{Department of Physics, Harvard University, 17 Oxford St, Cambridge, USA}
\emailAdd{randall@physics.harvard.edu}
\emailAdd{weishuangxu@g.harvard.edu}

\date{\today}

\abstract{
Low energy antideuteron detection presents a unique channel for indirect detection, targeting dark matter that annihilates into hadrons in a relatively background-free way. Since the idea was first proposed, many WIMP-type models have already been disfavored by direct detection experiments, and current constraints indicate that any thermal relic candidates likely annihilate through some hidden sector process. In this paper, we show that cosmic ray antideuteron detection experiments represent one of the best ways to search for hidden sector thermal relic dark matter, and in particular investigate a vector portal dark matter that annihilates via a massive dark photon. We find that the parameter space with thermal relic annihilation and $m_\chi > m_{A'} \gtrsim 20 \, \GeV$ is largely unconstrained, and near future antideuteron experiment \GAPS will be able to probe models in this space with $m_\chi \approx m_{A'}$ up to masses of $O(100\,\GeV)$. Specifically the dark matter models favored by the \textit{Fermi} Galactic center excess is expected to be detected or constrained at the $5(3)-\sigma$ level assuming a optimistic (conservative) propagation model. }

\begin{document} 

\maketitle
\flushbottom

\section{Introduction}
\label{sec:intro}

The search for dark matter (DM) is one of the defining challenges of contemporary particle physics, and although interesting anomalies have been seen throughout the multi-decade effort, no definitive evidence of a measurable particle interaction has yet been found. Though the current models in consideration cover a vast expanse of parameter space, the old idea of the Weakly Interacting Massive Particle (WIMP)-type thermal relic dark matter remains compelling, as a weak-scale $O(\pb)$ annihilation cross-section provides a natural origin for the present day dark matter abundance. However, direct detection experiments in particular are pushing constraints of DM-proton scattering to approaching ten orders of magnitude below that~\cite{Aalseth:2012if,Akerib:2016vxi,Agnese:2017njq,Amole:2017dex,Angle:2007uj}, and the scenario of simplistic weak-scale couplings to the visible sector is no longer viable.    

This is not to say that thermal relic dark matter scenario has been ruled out. Hidden sector dark matter, a class of models where DM annihilates to other dark sector particles, which in turn couple to the Standard Model (SM), is perhaps the most straightforward way to realize the WIMP scenario consistently with present-day constraints. These models can quite generally arise from the introduction of dark gauge fields. An annihilation cross section may be potentially detectable even though direct coupling to the light sector can be arbitrarily small. One particularly well-motivated type of DM, which we will consider in this work, is dark photon dark matter (DPDM), where DM particles are charged under a massive dark $U(1)$ and the dark photon kinetically mixes with the SM photon. Here, cascade annihilation is permitted as long as the dark photon is less massive than the dark matter particle. Dark photon dark matter and similar models have also been previously considered in e.g. Refs.~\cite{Cheung:2009qd,An:2014twa,Cohen:2010kn,Langacker:2008yv,Feng:2010gw,Co:2016akw,Hambye:2019dwd,Flambaum:2019cih,Tonnis:2019ppe}.

If the DM abundance is indeed set by thermal annihilation freeze-out, assuming no lighter particles in the dark sector, the energy density carried by the annihilated particles must be injected into either radiation, leptons, or a baryonic component of the universe. In the former case, measurements of $N_{eff}$ from the Cosmic Microwave Background (CMB) as probed by experiments like CMB-S4~\cite{Abazajian:2016yjj} are expected to detect or rule out this type of thermal relic. In the latter, if DM annihilates eventually into matter but direct coupling to the visible sector is to be sufficiently suppressed, indirect detection may be the best way to detect these models. 

Indirect detection efforts in the past years, measurements of CMB anisotropies~\cite{Aghanim:2018eyx}, and of gamma ray~\cite{Abdollahi:2017nat} and cosmic ray antiparticle~\cite{Aguilar:2016kjl} fluxes, have provided insightful constraints and even some intriguing excesses that could possibly point to a DM origin, such as the Galactic center excess (GCE) seen by the \textit{Fermi} collaboration~\cite{TheFermi-LAT:2017vmf,Brandt:2015ula,Lee:2015fea,Bartels:2015aea,Daylan:2014rsa,Gordon:2013vta,Hooper:2010mq,Abazajian:2012pn} and the anitproton excess seen by the Alpha Magnetic Spectrometer (\textit{AMS-02})~\cite{Cui:2016ppb,Cholis:2019ejx, Boudaud:2019efq}. Indeed, 10-40 GeV models of DPDM has been shown to be favored by the GCE if a modified NFW may be assumed for the galactic halo~\cite{Martin:2014sxa}. The origins of these excesses are still debated, the uncertainty of which dominantly arises from their relatively high astrophysical backgrounds. This necessitates accurate priors on poorly-understood quantities, such as secondary antiproton fluxes or distributions of millisecond pulsars, to identify a definitive excess. 

Cosmic ray antideuterons have not yet been confirmed by current running experiments, but they are attractive as a DM annihilation channel precisely because of their astrophysical rarity, particularly at low energy. They are generically produced by kinematically viable annihilating DM models ($m_\chi > m_{\dbar}$) that couple to quarks, albeit typically with low branching ratios. This potential downside is counterbalanced by the low expected secondary flux particularly at low kinetic energies, which is so low that experiments searching for cosmic ray antideuterons are virtually background free, and signals from annihilating DM can easily be orders of magnitude larger~\cite{Baer:2005tw}. The General AntiParticle Spectrometer (\textit{GAPS})~\cite{Ong:2017szd}, scheduled to launch in 2020-2021, is a balloon borne experiment using novel detection methods involving the decay of exotic atoms, designed particularly to search for antideuterons at low kinetic energies.

In conjunction with the \textit{AMS-02} experiment currently in operation, \GAPS has exciting potential to probe compelling DM models that may be difficult to study through other channels~\cite{Aramaki:2015pii,Dal:2014nda,Fornengo:2013osa,Ibarra:2012cc,Brauninger:2009pe,Cui:2010ud}. Indeed, the potential of detecting DM via cosmic ray antihelium has entered the discussion, motivated in part by eight antihelium events recorded recently by \AMS~\cite{Olivia:2019dbar}--though they are difficult to explain with known astrophysical or dark processes consistent with existing bounds~\cite{Blum:2017qnn,Korsmeier:2017xzj,Poulin:2018wzu}.   

In this article we describe dark photon dark matter and assess the prospects of detecting this type of dark matter via cosmic ray antideuterons with ongoing and near future experiments. In section~\ref{sec:model} we describe the details of the model; in section~\ref{sec:const} we discuss constraints on the parameter space from previous experiments; in section~\ref{sec:inject} we study the injection of antideuterons from DM annihilation into the galactic halo, and in section~\ref{sec:prop} the propagation of injected antideuteron through the interstellar medium and the solar environment. In section~\ref{sec:detect} we  present the main results for this paper: the expected top-of-atmosphere antideuteron flux for DPDM annihilation and the prospects of detection via \GAPS and \textit{AMS-02}. In section~\ref{sec:antinuc}  we discuss briefly the prospects of detecting antihelium from annihilated DPDM; in section~\ref{sec:concl} we summarize and conclude.


\section{Dark Photon Dark Matter}
\label{sec:model}

In this work, we consider a simple vector portal model in which DM consists of one flavor of Dirac fermion $(\chi, \bar\chi)$ with mass $m_\chi$, charged under a dark $U(1)$,   

\be 
\L \supset - g_D \bar\chi \gamma^\mu A'_\mu \chi + \frac{1}{2} m_\chi^2 \bar \chi \chi + \frac{1}{2} m_{A'}^2 A'_\mu A'^{\mu}.
\ee
The dark photon has St\"uckelberg  mass $m_{A'}$  and kinetically mixes with the SM photon, governed by parameter $\epsilon$,
\be \L \supset    - \frac{1}{4} F_{\mu\nu} F^{\mu\nu}  - \frac{1}{4} F'_{\mu\nu} F'^{\mu\nu} - \frac{\epsilon}{2}F'_{\mu\nu}F^{\mu\nu}.
\ee
Here $A'_\mu$ and $F'_{\mu\nu}$ are the dark counterparts to  the SM $U(1)$ gauge terms $A_\mu$ and $F_{\mu\nu}$.  The mixing between the dark and SM photons is expected for any separate U(1) symmetry in the universe~\cite{Holdom:1985ag}, so it is not an additional requirement of the model.  This mixing is diagonalized under the rotation 
\be 
A_\mu \to A_\mu + \epsilon A'_\mu  \qquad A'_\mu \to A'_\mu, 
\ee
which then manifests as an interaction between the dark photon and standard model fermions with coupling $\epsilon e_f$, where $e_f$ is the fermion's SM electric charge,
\be
\L \supset  - \epsilon e_f \bar\psi_f \gamma^\mu A'_\mu \psi_f.
\ee

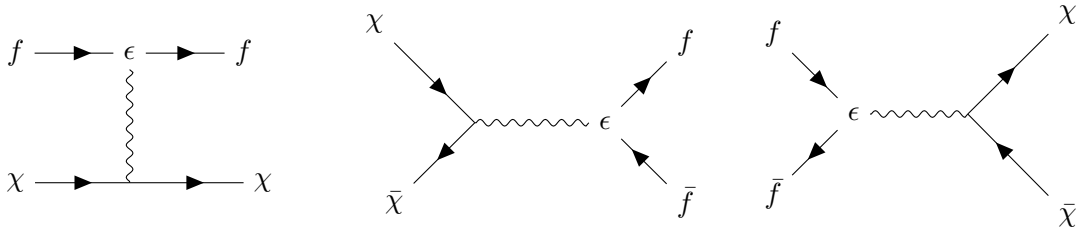
\begin{figure}
\begin{subfigure}{0.32\textwidth}
\begin{center} 
\begin{tikzpicture}
  \begin{feynman}
    \vertex (a){$\chi$};
    \vertex [right=of a] (b);
    \vertex [right=of b] (c){$\chi$};
    \vertex [above=of b] (d){$\epsilon$};
    \vertex [right=of d] (e){$f$};
    \vertex [left=of d] (f){$f$};
 
    \diagram* {
      (a) -- [fermion] (b),
      (c) -- [anti fermion] (b),
      (b) -- [boson] (d),
      (d) -- [fermion] (e),
      (d) -- [anti fermion] (f),
    };
  \end{feynman}
\end{tikzpicture}
\end{center}
\end{subfigure}
~
\begin{subfigure}{0.32\textwidth}
\begin{center} 
\begin{tikzpicture}
  \begin{feynman}
    \vertex (a){$\bar \chi$};
    \vertex [above right=of a] (b);
    \vertex [above left =of b] (f1){$\chi$};
    \vertex [right=of b] (c){$\epsilon$};
    \vertex [above right=of c] (f2){$f$};
    \vertex [below right=of c] (f3){$\bar f$};
    \diagram* {
      (b) -- [fermion] (a),
      (b) -- [anti fermion] (f1),
      (b) -- [boson] (c),
      (c) -- [fermion] (f2),
      (c) -- [anti fermion] (f3),
    };
  \end{feynman}
\end{tikzpicture}
\end{center}
\end{subfigure}
~
\begin{subfigure}{0.32\textwidth}
\begin{center} 
\begin{tikzpicture}
  \begin{feynman}
    \vertex (a){$\bar f$};
    \vertex [above right=of a] (b){$\epsilon$};
    \vertex [above left =of b] (f1){$f$};
    \vertex [right=of b] (c);
    \vertex [above right=of c] (f2){$\chi$};
    \vertex [below right=of c] (f3){$\bar\chi$};
    \diagram* {
      (b) -- [fermion] (a),
      (b) -- [anti fermion] (f1),
      (b) -- [boson] (c),
      (c) -- [fermion] (f2),
      (c) -- [anti fermion] (f3),
    };
  \end{feynman}
\end{tikzpicture}
\end{center}
\end{subfigure}
\caption{Relevant Feynman diagrams for SM-DM interactions: elastic scattering (left), $s$-channel annihilation of DM to SM (middle), $s$-channel production of DM from SM (right).  }
\label{fig:feyn_scatt}
\end{figure}

As shown in Fig.~\ref{fig:feyn_scatt}, the elastic scattering of $\chi, \bar\chi$ with SM particles, the $s$-channel direct annihilation of DM particles into the SM, as well as production of dark sector particles via collisions are quadratically dependent on $\epsilon$. If the mixing is small, the processes probed by direct detection and collider searches are both suppressed and occur inefficiently. Conversely, if annihilation occurred only along this channel, such as would be the case if $m_\chi < m_{A'}$,  cross sections allowed by existing direct detection constraints would be generically unobservable by indirect experiments and the thermal relic scenario would likewise be ruled out without further model building. 

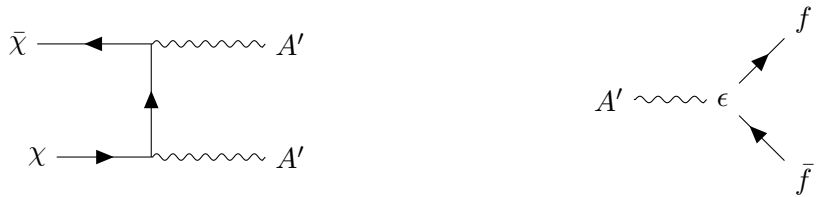
\begin{figure}
    \centering
\begin{subfigure}{0.45\textwidth}
\begin{center} 
\begin{tikzpicture}
  \begin{feynman}
    \vertex (a){$\chi$};
    \vertex [right=of a] (b);
    \vertex [above=of b] (c);
    \vertex [left=of c] (d){$\bar \chi$};
    \vertex [right=of c] (e){$A'$};
    \vertex [right=of b] (f){$A'$};
    \diagram* {
      (a) -- [fermion] (b),
      (b) -- [fermion] (c),
      (c) -- [fermion] (d),
      (c) -- [boson] (e),
      (b) -- [boson] (f),
    };
  \end{feynman}
\end{tikzpicture}
\end{center}
\end{subfigure}
~
\begin{subfigure}{0.45\textwidth}
\begin{center} 
\begin{tikzpicture}
  \begin{feynman}
    \vertex (a){$A'$};
    \vertex [right=of a] (b){$\epsilon$};
    \vertex [above right=of b] (f1){$f$};
    \vertex [below right=of b] (f2){$\bar f$};
    \diagram* {
      (a) -- [boson] (b),
      (b) -- [fermion] (f1),
      (b) -- [anti fermion] (f2),
    };
  \end{feynman}
\end{tikzpicture}
\end{center}
\end{subfigure}
\caption{Feynman diagrams for the cascade annihilation of dark matter into on-shell dark photons (left) and subsequent decay of dark photons to SM fermions (right).  }
\label{fig:feyn_ann}
\end{figure}

 However in the case that  $m_\chi > m_{A'}$, the $t$-channel annihilation of DM into on-shell dark photons, as shown in Fig.~\ref{fig:feyn_ann}, is also kinematically allowed. These dark photons are allowed to propagate and decay into SM particles over a longer characteristic time, and in this way the annihilation into the light sector circumvents the suppression of the mixing. The annihilation of DM, dominated by this process, is therefore much more efficient than elastic scattering or collider production. This type of hidden sector mechanism thus allows in a natural way the realization of a thermal relic abundance in light of increasingly stringent direct detection constraints, and it naturally follows that indirect detection is the most advantageous way to search for this type of dark matter. 

\section{Constraints from Previous Experiments}
\label{sec:const}

\begin{figure}
    \centering
    \begin{subfigure}[t]{0.5\textwidth}
        \centering
        \includegraphics[width=\linewidth]{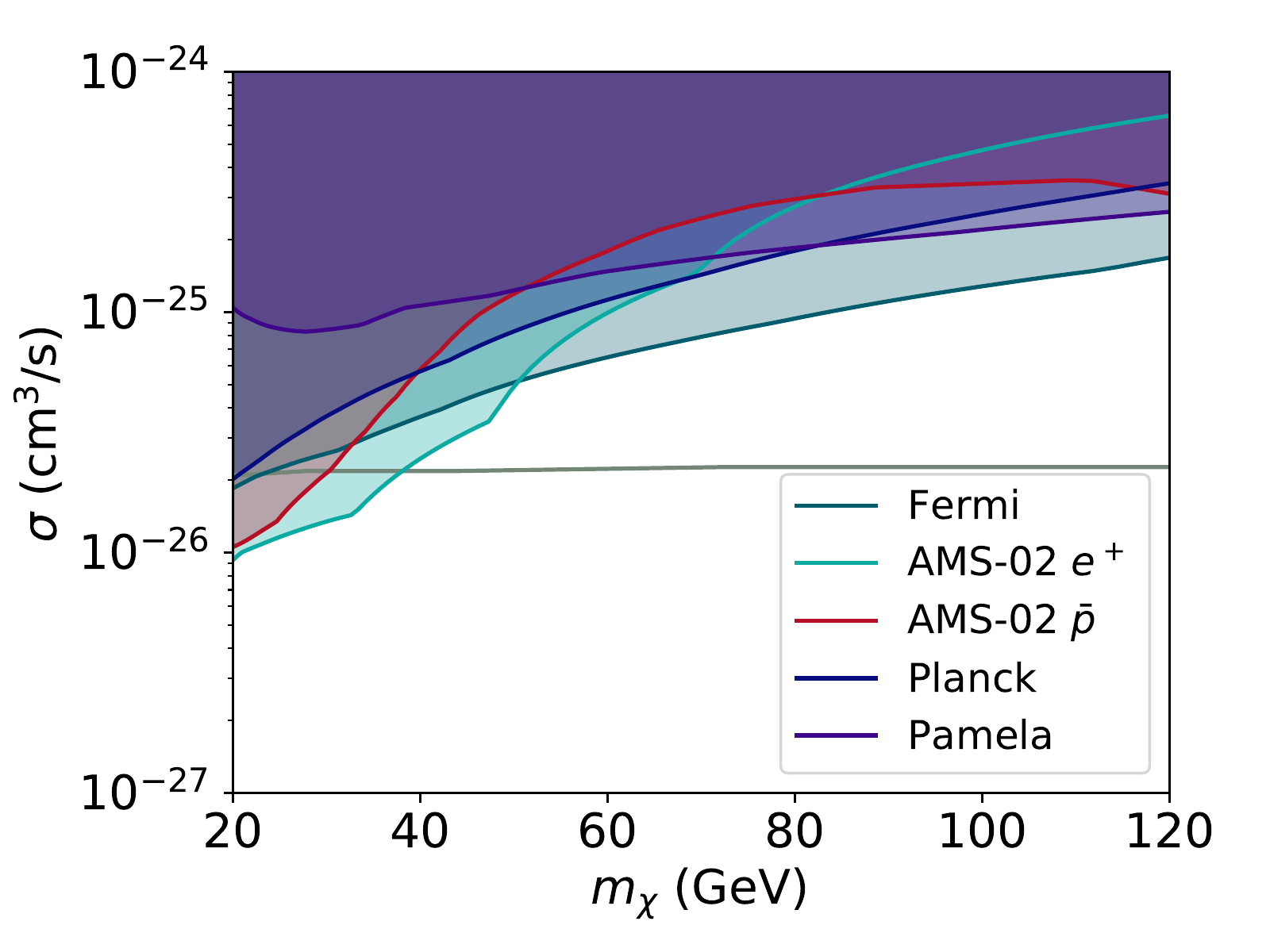}
        \caption{Constraints on annihilating cross section from  indirect detection experiments \textit{Planck}~\cite{Ade:2015xua}, \AMS positrons~\cite{Aguilar:2014mma} and antiprotons~\cite{Aguilar:2016kjl}, \textit{PAMELA}~\cite{Adriani:2013uda} and \textit{Fermi-LAT}~\cite{Ackermann:2015zua}. The grey line denotes the thermal relic cross section that sources the present-day DM abundance. Here, $m_{A'}/m_\chi$ is taken to be 1/2.}
        \label{fig:indt_const}
    \end{subfigure}%
    ~ 
    \begin{subfigure}[t]{0.5\textwidth}
        \centering
        \includegraphics[width=\linewidth]{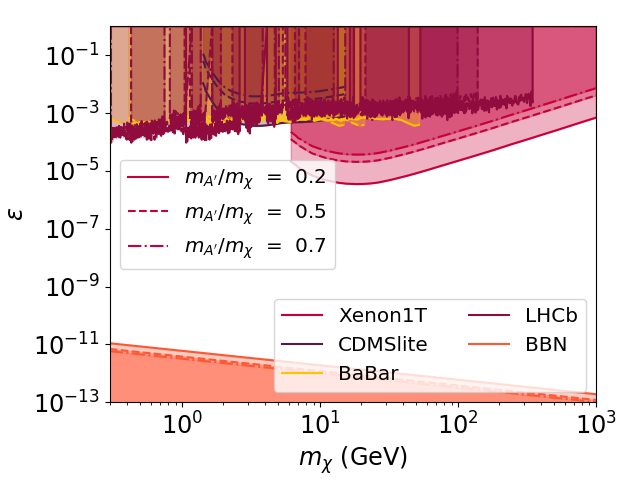}
    \caption{Constraints on kinetic mixing $\epsilon$ from direct detection results from \textit{Xenon1T}~\cite{Angle:2007uj}, \textit{CDMSlite}~\cite{Agnese:2018gze}, collider experiments \textit{LHCb}~\cite{Aaij:2019bvg}, \textit{BaBar}~\cite{Lees:2014xha}, and BBN. Here, $\langle \sigma v \rangle_{ann}$ is taken to be 1 pb.}
        \label{fig:epsconst}
    \end{subfigure}
    \caption{Constraints on dark photon dark matter parameter space from previous experiments.}
    \label{fig:exp_consts}
\end{figure}

The most direct constraints on the parameter space of this model derive from previous indirect detection experiments and limits on CMB energy injection or fluxes of other cosmic ray species. These constraints are analyzed in~\cite{Elor:2015bho,Cui:2018klo,Fornengo:2013xda} and are shown for the specific case of DPDM in Fig.~\ref{fig:indt_const}. Here, we have assumed $m_{A'}/ m_{\chi} = 0.5$, though the constraints are not particularly sensitive to this mass ratio. As shown, DPDM annihilating at thermal relic cross sections is ruled out for $m_\chi \lesssim 40 \,\GeV$. Constraints become significantly weaker for heavier particle masses. 

While the signatures of DPDM are most readily seen via indirect detection, the annihilation channel is ill-equipped to probe the kinetic mixing precisely because it is insensitive to the suppressed coupling. In this case, direct detection and collider searches potentially offer the means to measure and constrain the mixing parameter $\epsilon$ once the annihilation cross section is measured by indirect experiments. Assuming that $m_\chi > m_{A'}$ and the two-step annihilation process is allowed to occur, the cross section is given by 
\be 
\sigma_{ann} = \frac{\pi \alpha_D}{m_\chi^2} \sqrt{1-\frac{m_{A'}^2}{m_\chi^2}},
\ee
where $\alpha_D = g_D^2/4\pi$. The DM-proton elastic scattering however, is quadratically suppressed by $\epsilon$, 
\be
\sigma_{\chi p} = \frac{4 \alpha_D e^2 \epsilon^2}{m^4_{A'}} \frac{m_\chi^2 m_p^2}{(m_\chi + m_p)^2 }.
\ee

Taking a fiducial thermal relic cross section $\sigma_{ann} = 1 \pb$, constraints on $\sigma_{\chi p}$ as obtained from direct detection translate to constraints on $\epsilon^2$ given choices of $m_\chi, m_{A'}$. Constraints on dark photon couplings have also been measured directly by collider experiments, and are independent of thermal relic assumptions. 

On the other hand, lower-bound constraints for $\epsilon$ are set such that the mixing and decay of dark photons into the light sector occur sufficiently quickly so as not to interfere with Big Bang neucleosynthesis (BBN), 
\be 
\tau_{A'} \sim \left(e^2 \epsilon^2 m_{A'} \right)^{-1} < t_{BBN} \sim 1 \,\mathrm{s}.
\ee

A summary of these constraints, which yield both upper- and lower-bound on $\epsilon$, are given in Fig~\ref{fig:epsconst} for various choices of $m_{A'} < m_\chi$. These constraints become less stringent, particularly for direct detection contours, as $m_{A'}$ approaches $m_\chi$, but in all cases there are roughly six orders of magnitude of unconstrained choices of $\epsilon$, demonstrating significant remaining freedom in the parameter space of DPDM.

\section{Coalescence and Injection into the Halo}
\label{sec:inject}

While compared to heavier nuclei a substantial amount of data on antideuteron production at colliders has been recorded~\cite{Acharya:2017fvb,Asner:2006pw,Lees:2014iub},  the process of antideuteron formation is not well known, and the common practice is a phenomenological parameterization of coalescence calibrated to match experimental results. The dependence of this formation process on the content of background particles and on the center of mass energy is still unknown and constitutes an active field of study.  Thus, the coalescence model is an important source of uncertainty in antideuteron predictions, and best-fit parameters to different datasets may induce a difference in production efficiency by a factor of a few.

For this work, we will use the step-function coalescence criterion: the approximation that a nucleus is formed if the constituent nucleons are formed within a sphere in momentum space of diameter $p_0$, in the center-of-mass frame~\cite{Aramaki:2015pii}.  In the case of antideuteron production, a  $\dbar$ is formed if a produced $\pbar, \nbar$ pair have a center-of-mass momentum separation less than $p_0$.  

\begin{figure}
    \centering
    \includegraphics[width=0.7\linewidth]{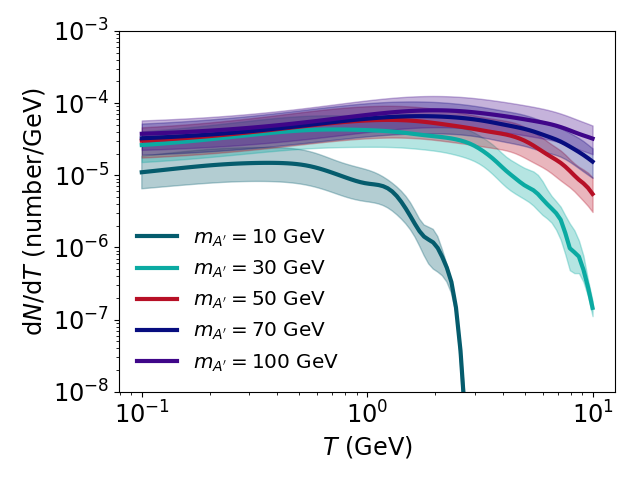}
    \caption{Injection spectra obtained \textbf{with {\tt Pythia 8}}  for various choices of dark photon mass $m_{A'}$. The shaded region shows a range of coalescence momenta $p_0 = 150 \pm 25$ MeV; the wiggles in the spectra are numerical artifacts.}
    \label{fig:mA_injspec}
\end{figure}

\begin{figure}
    \centering
    \includegraphics[width=0.7\linewidth]{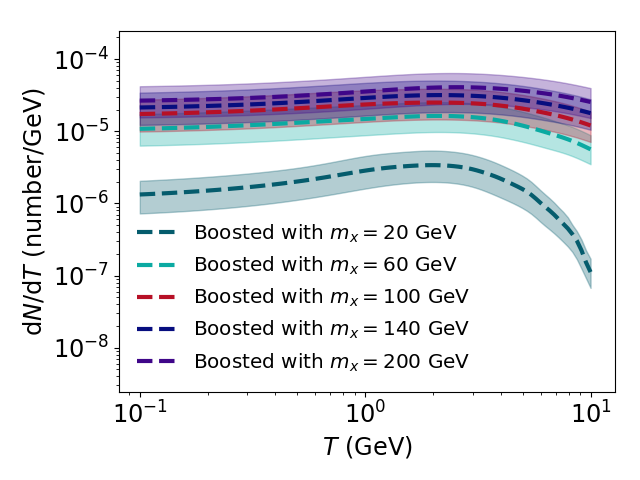}
    \caption{Similar to Fig.~\ref{fig:mA_injspec}, but isotropically boosted to the dark matter (halo) rest frame assuming $m_\chi = 2 m_A'$.}
    \label{fig:injspec}
\end{figure}

The coalescence momentum is set to match anti-nuclei production observed in collider experiments~\cite{Ibarra:2012cc,Dal:2014nda} and the best-fit values for modeling antideuteron production using {\tt Pythia 8} listed in Table~\ref{tab:coals}.   The coalescence criterion is then applied on a per-event basis in a Monte Carlo analysis \textbf{with {\tt Pythia}} to determine the injection spectra from the decay $A' \to  \dbar X$ in the rest frame of the dark photon, shown for example dark photon masses in Fig.~\ref{fig:mA_injspec}. The branching ratios of the dark photon are inherited from the couplings of the SM photon with which it mixes.  Assuming the dark matter annihilations occur in the rest frame of the galaxy, and the dark photons are produced isotropically, the boosted injection spectrum into the galactic halo per dark matter annihilation is given for sample dark matter masses in Fig.~\ref{fig:injspec}.  Experimental searches focused on low kinetic energy events, where the background is most subdominant, will be most sensitive to models with $m_\chi \approx m_{A'}$, though models with extremely close masses may be less generic.  We choose as a benchmark a model with $m_\chi = 50\, \GeV$, $m_{A'} = 30\,\GeV$, and investigate the detection prospects for the  satellite experiments \textit{GAPS} and \textit{AMS-02}.

\begin{table}[t]
    \centering
    \begin{tabular}{c|c  c c c}
         $\sqrt{s}$ [GeV] &  9.46  & 10.58  & 53  & 91.19   \\
         \hline 
         Experiment & CLEO ($\Upsilon$ resonance)  & BaBar ($e^+ e^-$) & ISR ($pp$) & ALEPH ($Z$ resonance) \\
         $p_0$ [MeV] &  133 &  135 & 152 & 192
    \end{tabular}
\caption{Coalescence momenta for antideuteron formation as fit using {\tt Pythia 8} from experimental data. Taken from~\cite{Ibarra:2012cc,Dal:2014nda}}
    \label{tab:coals}
\end{table}

\section{Propagation through the Galaxy and Heliosphere}
\label{sec:prop}

After coalesced antideuterons are injected into the galaxy, the transport of the cosmic ray to the detector is described by two components: the propagation through the interstellar medium (ISM) to the Solar System, and the propagation through the heliosphere to the top of the Earth's atmosphere (TOA).    Galactic transport in particular constitutes a significant source of uncertainty in this analysis, since as with the coalescence picture, parameters of the transport model are phenomenological.  They are calibrated to match cosmic ray fluxes measured by experiments, and different supported choices for parameters may result in differences in the derived flux of close to an order of magnitude~\cite{Aramaki:2015pii,Fornengo:2013osa}.   

The galactic propagation of cosmic ray antinuclei is described by a two-zone diffusion model~\cite{Maurin:2001sj,Donato:2001eq,Maurin:2002hw}:  production and transport is assumed to occur within a cylinder axis-aligned with the axis of the galactic plane, of radius $R$ and half-height $L$ above and below the galactic plane. The ISM where disintegration of the antinuclei can occur is confined to a thin disk of half-height $h$ above and below this plane.  As done in Refs.~\cite{Maurin:2001sj,Donato:2001eq,Maurin:2002hw}, we set $R = 20 \, \kpc,  h = 100\, \pc$, and take $L$ as a transport parameter.   The number density of antinuclei $n(r,z,T)$ at position $(r,z)$ and kinetic energy $T$ is governed by the transport equation
\be
-\vec \nabla \cdot [ K (r,z,T) \vec \nabla n(r,z,T) + \vec V_c n(r,z,T)] + 2h \delta(z) \Gamma^{\dbar}_{dst} n(r,z,T) = Q^{\dbar}(r,z,T),
\ee
whose terms we will now detail below. $Q^{\dbar}(r,z,T)$ is the source term of antideuteron production, which in the case of DM annihilation is quadratically dependent on the halo profile:
\be
Q^{\dbar}(r,z,T) =  \frac{1}{2} \langle \sigma v \rangle \frac{dN_{\dbar}}{dT} \left( \frac{\rho(r,z)}{m_\chi} \right)^2.
\ee
Here, $dN_{\dbar}/dT$ is the injection spectrum computed in Section~\ref{sec:inject}, and $\langle \sigma v \rangle$ is the thermal average annihilation cross section. To realize  thermal relic dark matter, we take $\langle \sigma v \rangle = 1\pb$. We assume a spherical NFW profile for the galactic halo 
\be \label{eq:nfw}
\rho (r,z) =  \frac{ \rho_\odot r_\odot (1 + r_\odot/r_s)^2}{ \sqrt{r^2 + z^2}( 1 + \sqrt{r^2+z^2}/r_s )^2 }, \qquad r_s = 20\,\kpc.
\ee
The final observed flux has some dependence on the choice of halo profile, but the sensitivity is small compared to uncertainties in transport parameters. To demonstrate this we will also consider a spherical Einasto profile, 
\be\label{eq:ein}
\rho (r,z) = \rho_\odot \exp \left(- \frac{2}{\alpha}\left[ \left(\frac{\sqrt{r^2+z^2}}{r_s} \right)^\alpha - \left(\frac{r_\odot}{r_s}\right)^\alpha \right] \right), \qquad r_s = 20 \, \kpc, \quad \alpha =0.17, 
\ee
and a cored isothermal profile
\be \label{eq:iso}
\rho (r,z) =  \frac{ \rho_\odot (1+ r_\odot^2/r_s^2)}{1 + (r^2 + z^2)/r_s^2}, \qquad r_s = 5\,\kpc,
\ee 
all of which are fixed such that the density is $\rho_\odot = 0.39 \, \GeV/\cm^{3}$. at $r_\odot = 8.5 \, \kpc$. We find that overall the Einasto profile gives a slightly optimistic flux compared to the NFW case, and the isothermal gives a slightly conservative one.

The annihilation of antideuterons in the ISM is governed by the term 
\be
\Gamma^{\dbar}_{dst} = (n_H + 4^{2/3}n_{He})v_{\dbar} \sigma_{ine}^{\dbar p \to X },
\ee
where $n_H = 1 \, \cm^{-3}$, $n_{He} = 0.1 \, \cm^{-3}$ are the number densities of hydrogen and helium in the ISM, and $v_{\dbar}$ is the velocity of the cosmic ray $\dbar$. The inelastic cross section $\sigma^{\dbar p \to X}_{ine}$ has not been measured, so we will follow the approximation scheme developed in~\cite{Brauninger:2009pe} using measured values for the charge conjugate processes $\sigma^{d\pbar}$.

Finally, the first term governs the diffusion and convection of the cosmic ray density. The convection wind is taken to be constant in magnitude pointing away from the galactic plane, $\vec V_c = \mathrm{sgn}(z) V_c \hat z$, and the diffusion coefficient is parameterized as 
\be
K(r,z, T) = v_{\dbar} K_0 \left( \frac{\mathcal{R}}{1 \, \mathrm{GV}}\right)^\delta,
\ee
where $\mathcal{R} = p/(Ze)$ is the cosmic ray rigidity in units of Gigavolts. Thus, the model for galactic transport is encoded in the choice of parameters $\{ L, V_c, K_0, \delta \}$, which are in turn determined by boron-to-carbon (B/C) cosmic ray ratios~\cite{2011A&A...526A.101P,2012A&A...544A..16T,Donato:2003xg} as so-called MIN, MED, and MAX propagation models. The first of these has recently become disfavored by more recent measurements ~\cite{Adriani:2013uda,Trotta:2010mx} so we will only consider the MED and MAX scenarios in this work. These parameter choices are independent of cosmic ray species and are tabulated in Table~\ref{tab:galprop}.

\begin{table}[]
    \centering
    \begin{tabular}{c|c c c c}
         &  $L \,[\kpc]$ & $V_c \,[\km/\s]$  & $K_0 \, [\kpc^2/\Myr]$  & $\delta$  \\
         \hline
        MED & 4 & 12 & 0.0112 & 0.70\\
        MAX & 15 & 5 & 0.0765 & 0.46\\
    \end{tabular}
    \caption{Galactic transport model parameters for the MED and MAX scenarios.}
    \label{tab:galprop}
\end{table}

The galactic propagation and dependencies on transport models can be encoded in an function $R^{\dbar}(T)$~\cite{Fornengo:2013osa}, which is independent of the choice of halo profile and DM model, for all annihilating DM scenarios. The interstellar flux can then be written as 
\be
\phi^{\dbar}_{ISM}(r,z,T) = \frac{v_{\dbar}}{8\pi} \langle \sigma v \rangle \frac{dN_{\dbar}}{dT} \left(\frac{\rho(r,z)}{m_\chi}\right)^2 R^{\dbar}(T), 
\ee
and we solve for this propagation function numerically in a Bessel expansion. We note that more sophisticated propagation models have been developed~\cite{Genolini:2019ewc,Reinert:2017aga} and reserve the inclusion of additional astrophysical effects for future work. The transport of cosmic rays in the solar environment, on the other hand, is governed by the modulation and structure of the solar magnetic field. We use the force field approximation~\cite{Gleeson:1968zza} to model this effect, writing the observed antideuteron flux at the top of the Earth's atmosphere as
\be
\phi^{\dbar}_{TOA}(T) = \phi^{\dbar}_{ISM}(r_\odot, 0, T') \frac{T^2 + 2m_{\dbar}T}{T'^2 + 2m_{\dbar}T'}, \qquad T' = T + |Z|e \Phi ,
\ee
and we take $\Phi = 500 \, \mathrm{MV}$, which reproduces the correct energy shift on average. It is shown in Ref~\cite{Fornengo:2013osa} sensitivity of the predicted flux to this approximation is subdominant to the uncertainties in coalescence momentum and galactic propagation parameters.

\section{Detection Prospects}
\label{sec:detect}

We consider experimental efforts to detect cosmic ray antideuterons, and specifically the current ongoing experiment \AMS and the near future experiment \textit{GAPS,} and evaluate the prospects of detecting antideuterons from DPDM annihilation.  Identification of antideuteron particles by cosmic ray detectors consists of three main components: measurement of the particles' charge, sign of charge, and mass. The last measurement is the most difficult. Due to their abundance, a major source of detection uncertainty derives from mislabeling of antiproton cosmic rays. 

\AMS is a general high-energy cosmic ray experiment that centers around a 0.14 T permanent magnet~\cite{Ting:2013sea,Aramaki:2015pii}. A silicon tracker, time-of-flight detector, transition radiation detector, and ring imaging Cherenkov detector each measure the charge of the cosmic ray, while curvature of the trajectory induced by the magnet indicates the sign of the charge. The mass is derived from measurements of rigidity and charge. For the purposes of antideuterons, the time-of-flight detector was planned to measure low energy particles with a sensitivity of
\be
\phi^{\dbar}_{\mathrm{AMS-02,\, low}} = 2.1 \times 10^{-6} \,  [\mtr^2 \, \s \, \sr \, \GeV]^{-1} \qquad 
T \in [0.2, 0.8] \,  \GeV, 
\ee
while the ring imaging Cherenkov detector was planned to measure particles with higher kinetic energy at a sensitivity of 
\be
\phi^{\dbar}_{\mathrm{AMS-02,\, high}} = 2.1 \times 10^{-6} \,  [\mtr^2 \, \s \, \sr \, \GeV]^{-1} \qquad 
T \in [2.2, 4.2] \,  \GeV, 
\ee
both for a five year observation period. However, this calculation is based on the originally planned 0.86 T super-conducting magnet. Actual sensitivities based on the magnet in use are not known, though it is safe to assume the reported numbers are an upper-bound. 

The \GAPS experiment on the other hand is focused specifically on antiparticle cosmic rays at low kinetic energies, exploring a parameter space complementary to \AMS~\cite{Aramaki:2015laa,Aramaki:2015pii}. The instrument contains a time-of-flight detector and layers of semiconducting targets. Antiparticles will be trapped to form short-lived exotic atoms which then decay. For the anticipated three flights totaling a period of 105 days, \GAPS reports a projected sensitivity of 
\be
\phi^{\dbar}_{\mathrm{GAPS}} =  2\times 10^{-6} \,  [\mtr^2 \, \s \, \sr \, \GeV]^{-1} \qquad 
T \in [0.05, 0.25] \,  \GeV.
\ee
We include additionally the upper bound of antideuteron fluxes reported by the Balloon-borne Experiment with Superconducting Spectrometer (\textit{BESS}) experiment for reference~\cite{Fuke:2005it,Aramaki:2015pii} 
\be
\phi^{\dbar}_{\mathrm{BESS}} <  1.8 \times 10^{-4} \,  [\mtr^2 \, \s \, \sr \, \GeV]^{-1} \qquad 
T \in [0.17, 1.15] \,  \GeV.
\ee

The production of secondary antideuterons, from the spallation of the ISM from cosmic rays (and subdominantly from supernovae remnants), occurs at a higher threshold energy  of 17$m_p$ (as opposed to 7$m_p$ for secondary antiprotons). As a result, the secondary antideuteron spectrum is highly suppressed at low kinetic energies; in contrast, dark matter particles annihilate effectively at rest, dominating the background in this regime, where experiments like \GAPS are most sensitive. In this work we take the computed secondary flux from Ref.~\cite{Herms:2016vop} as our expected background.

\begin{figure}
    \centering
    \includegraphics[width=0.7\linewidth]{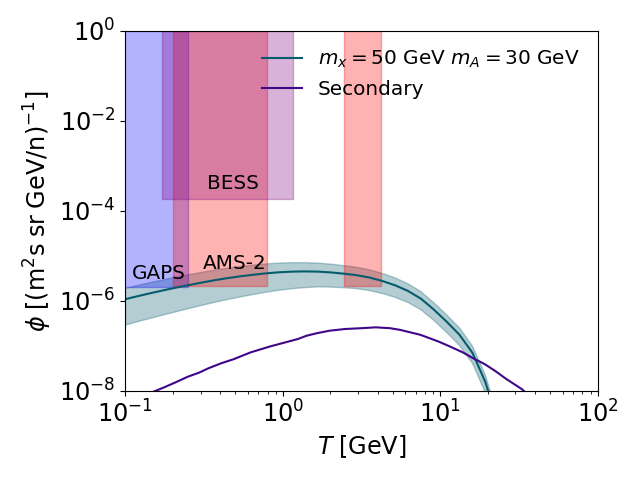}
    \caption{The antideuteron flux spectrum of a thermal relic dark photon dark matter with $m_\chi = 50\,\GeV$ and $m_{A'} = 30\,\GeV$. We take here $p_0 = 150 \, \MeV$ and assume an NFW profile for the halo; the shaded region represents uncertainty between MED and MAX propagation models. Reported sensitivities of \GAPS~\cite{Aramaki:2015laa}, \textit{BESS}~\cite{Fuke:2005it}, and \AMS~\cite{Ting:2013sea} (with superconducting magnet), as well as the projected astrophysical antideuteron flux~\cite{Herms:2016vop} are shown for reference.}
    \label{fig:flux}
\end{figure}

\begin{figure}
    \centering
    \begin{subfigure}[t]{0.5\textwidth}
        \centering
        \includegraphics[width=\linewidth]{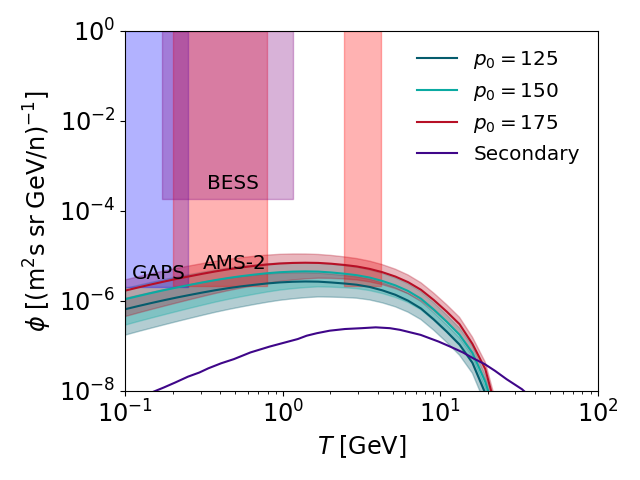}
    \end{subfigure}%
    ~ 
    \begin{subfigure}[t]{0.5\textwidth}
        \centering
        \includegraphics[width=\linewidth]{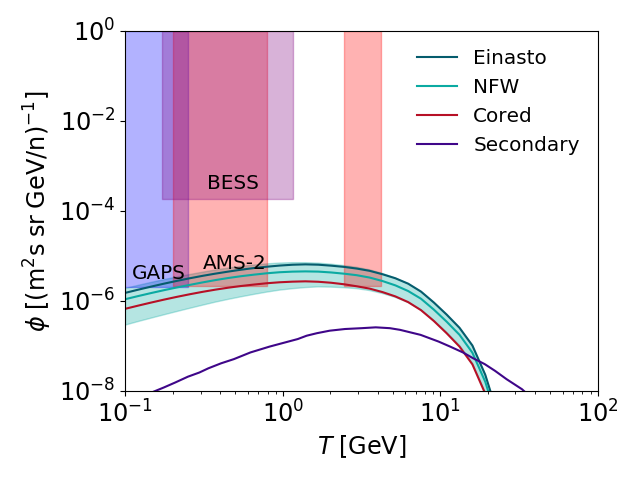}
    \end{subfigure}
    \caption{The sensitivity of expected antideuteron flux ($m_\chi = 50\, \GeV, m_{A'} = 30\, \GeV$) to variations in coalescence momenta (left) and halo profiles (right) used during injection and propagation steps. }
    \label{fig:flux_vary}
\end{figure}

For our benchmark model of $m_\chi = 50\, \GeV, m_{A'} = 30\, \GeV$, assuming a coalescence momentum of $p_0 = 150\, \MeV$ and an NFW halo profile, the projected antideuteron flux from a thermal relic annihilation is shown in Fig~\ref{fig:flux}. The shaded region denotes the uncertainty in propagation model. We see that the annihilation flux is several orders of magnitude above the expected secondary.  Fig~\ref{fig:flux_vary} shows the change in expected flux with variation of coalescence momenta (left) and halo profiles (as described in Eqs.~\ref{eq:nfw}~-~\ref{eq:iso}, right). The uncertainty in flux from these two sources are each less than but comparable to that from the MED and MAX propagation models, each inducing a change of $\sim 20\%$. Dependence on annihilation cross section is multiplicative and is not shown.

\begin{figure}
    \centering
    \begin{subfigure}[t]{0.49\textwidth}
        \centering
        \includegraphics[width=\linewidth]{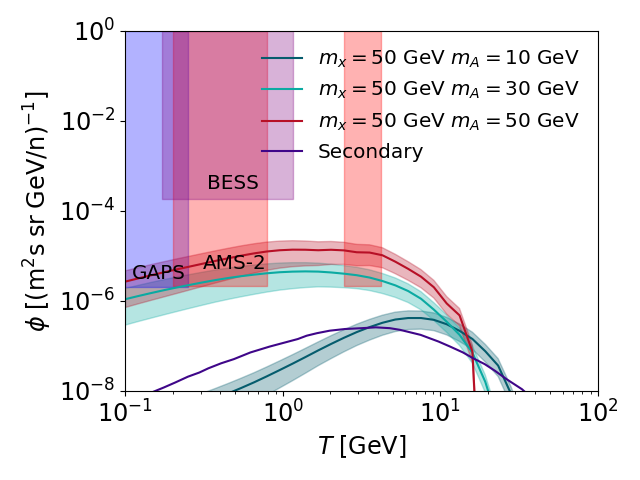}
    \end{subfigure}%
    ~ 
    \begin{subfigure}[t]{0.49\textwidth}
        \centering
        \includegraphics[width=\linewidth]{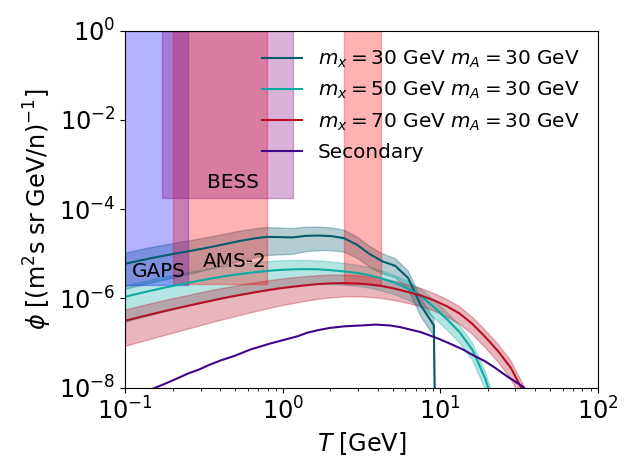}
    \end{subfigure}
    \caption{The dependence of expected antideuteron flux for various choices of $m_{A'}$ (left) and $m_\chi$ (right). As shown, the expected flux at low energies in particular is quite sensitive to the mass difference between these two particles, and the antideuteron injection is much less efficient for low dark photon masses.} 
    \label{fig:flux_mass_vary}
\end{figure}

While this is the benchmark mass model we have chosen, a large range of dark matter and dark photon masses are potentially accessible by antideuteron experiments. Fig~\ref{fig:flux_mass_vary} shows the change in flux spectrum for various dark photon masses, holding $m_\chi = 50\, \GeV$ (left), and various dark matter masses, holding $m_{A'}=30\, GeV$. As these experiments, \GAPS particularly, focuses on low kinetic energy cosmic rays, they are optimally suited for discovering moderately light $m_\chi \lesssim 100 \, GeV$ DM models where $m_{A'} \approx m_\chi$. 

To quantify the detectability of DPDM with the \GAPS detector, we evaluate the expected number of signal and background events seen by the experiment. This number is given by 
\be
N_{events} = \int dT \phi_{\dbar}(T) \tau (\Gamma_s(T) + \Gamma_a(T)),
\ee
where $\tau$ denotes the total flight time of 105 days, and $\Gamma_s$ and $\Gamma_a$ are taken from Ref.~\cite{Aramaki:2015laa} and represent the experimental reaches for stopped events and in-flight annihilation events respectively. 

The projected secondary flux for antideuterons yields an expected $b= 0.038$ background events for the entire flight time, and the experiment is thus virtually background-free. We take $N_{x\sigma}$ to be the critical number of Poissonian events necessary to make an $x \sigma$ detection:
\be \sum^{N_{3\sigma}-1}_{n=0} P(n,b) > 0.997, \qquad \sum^{N_{5\sigma}-1}_{n=0} P(n,b) > 0.9999994. \ee
We find that (with \GAPS) $N_{3\sigma} = 2$ events are sufficient for a $3\sigma$ detection, and $N_{5\sigma} = 4$ events are sufficient for a $5\sigma$ discovery.

\begin{figure}
    \centering
    \begin{subfigure}[t]{0.49\textwidth}
        \centering
        \includegraphics[width=\linewidth]{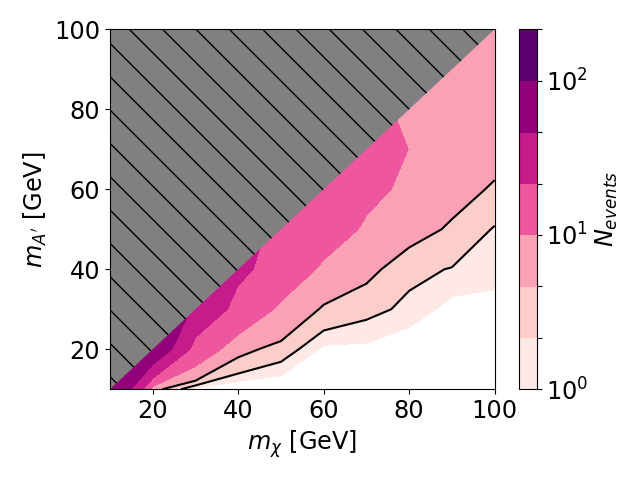}
    \end{subfigure}%
    ~ 
    \begin{subfigure}[t]{0.49\textwidth}
        \centering
        \includegraphics[width=\linewidth]{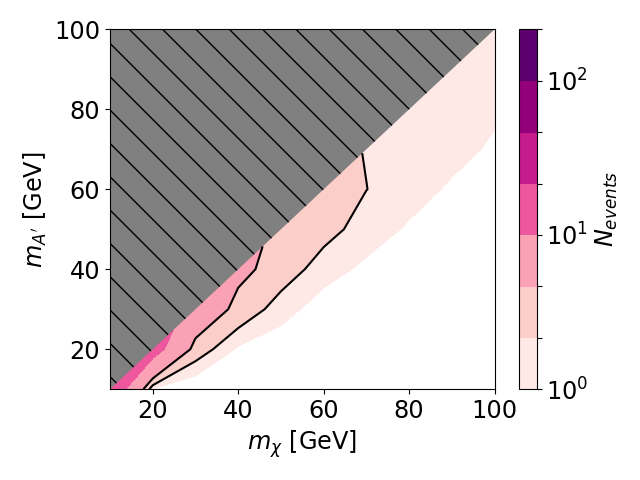}
    \end{subfigure}
    \caption{The expected number of events observable by \GAPS over the parameter space of DPDM model assuming MAX (left) and MED (right) cosmic ray propagation. A NFW profile as in Eq.~\ref{eq:nfw} and coalescence momentum of $p_0 = 150 \,\MeV$ are assumed. The contours represent $3\sigma$ ($N=2$) and $5\sigma$ ($N=4$) detection thresholds respectively. Our chosen benchmark model predicts a $>5\sigma$ detection in the most optimistic scenario and a $2-3\sigma$ detection in the most conservative.}
    \label{fig:Nevents}
\end{figure}

Fig.~\ref{fig:Nevents} shows the expected number of events seen by \GAPS for a range of models in the $m_{A'} - m_\chi$ parameter space, assuming MAX (left) and MED (right) propagation models. A thermal relic cross section of $\langle \sigma v \rangle  = 1 \text{ pb}$ is assumed, and coalescence and halo profiles are taken as in Fig.~\ref{fig:flux}, though we note that for $m_\chi \lesssim 40 \GeV$ this cross section is challenged by bounds discussed in Sec.~\ref{sec:const}.  The region enclosed by the black contours shows the mass ranges detectable by \GAPS at a $3-$ and $5\sigma$ level respectively. As shown, the measurability of a large portion of this parameter space depends greatly on the propagation model, but even in the least optimistic scenario $3\sigma$ measurements can be made of DM masses of up to $80 \GeV$ for $m_{A'}$ sufficiently close to $m_\chi$. 

In complementarity with other indirect searches, the excess of gamma rays in the Galactic center observed by \textit{Fermi-LAT} has been robustly established, but a consensus on its origins has yet to be reached~\cite{TheFermi-LAT:2017vmf,Brandt:2015ula,Lee:2015fea,Bartels:2015aea,Daylan:2014rsa,Gordon:2013vta,Hooper:2010mq,Abazajian:2012pn}. The excess of gamma ray flux can be modeled with dark matter that annihilates into quarks or leptons
\be
\phi^{\gamma} = \frac{\langle \sigma v \rangle}{ 16 \pi m^2_\chi} \frac{d N_\gamma}{ dT} \int_{los} \du l \; \rho^2 (r, z),
\ee
where the DM distribution in the line of sight is integrated over, and the measurement is in fact consistent with DM annihilating at a thermal relic cross section~\cite{Daylan:2014rsa,Hooper:2010mq}. The excess may also be attributed to unresolved point sources, most likely millisecond pulsars~\cite{Lee:2015fea,Bartels:2015aea}, and the preference of the signal to one scenario or the other is still hotly debated. Previous work has suggested that if the GCE observed by \textit{Fermi} indeed has dark matter origins then a vector portal DM model provides a good candidate for observed signal~\cite{Martin:2014sxa}.  For this analysis a steeper modified NFW is assumed for the galactic halo is assumed 
\be
\rho_\gamma (r,z) =  \frac{ \rho_\odot (1 + r_\odot/r_s)^2 (r_\odot/\sqrt{r^2 + z^2})^{\gamma} }{ ( 1 + \sqrt{r^2+z^2}/r_s )^{3-\gamma} }, \qquad r_s = 20\,\kpc \quad \gamma = 1.26.
\ee

\begin{figure}
    \centering
    \begin{subfigure}[t]{0.49\textwidth}
        \centering
        \includegraphics[width=\linewidth]{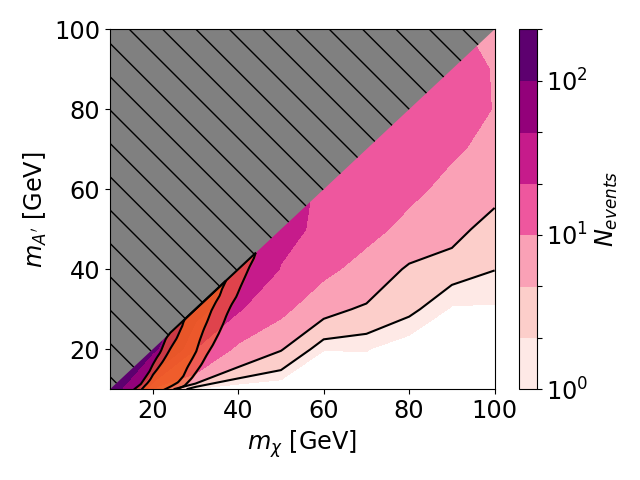}
    \end{subfigure}%
    ~ 
    \begin{subfigure}[t]{0.49\textwidth}
        \centering
        \includegraphics[width=\linewidth]{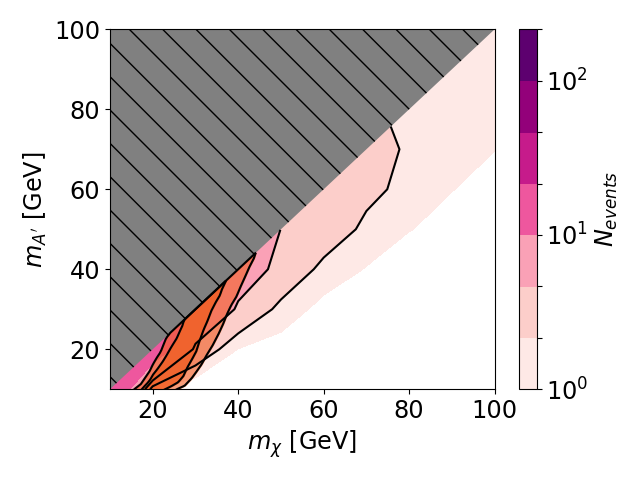}
    \end{subfigure}
    \caption{Similar to Fig.~\ref{fig:Nevents}, the expected number of events observable by \GAPS for the DPDM model in a modified NFW with $\gamma = 1.26$ assuming MAX (left) and MED (right) cosmic ray propagation. The region favored by the GCE measurement is highlighted and taken from~\cite{Martin:2014sxa}.}
    \label{fig:Nevents_gce}
\end{figure}

Fig~\ref{fig:Nevents_gce} shows the expected number of events detected by \GAPS for DPDM distributed in a modified NFW, with the region highlighted region being favored by the GCE at the $68\%$ and $95\%$ CL assuming MAX (left) and MED (right) propagation parameters. Thus, a DPDM model origin for the \textit{Fermi} Galactic center excess is expected to be detected or constrained by \GAPS at the $5\sigma$ level in the optimistic, or the $3\sigma$ level for the conservative case. 

\section{Antihelium flux}
\label{sec:antinuc}

In this section we discuss the potential for annihilating DPDM to generate antihelium events that could be seen by \GAPS or \AMS, in light of the recently reported eight antihelium events by \AMS~\cite{Olivia:2019dbar}, which consists of six $\bar{\mathrm{He}_3}$ and two  $\bar{\mathrm{He}_4}$ events. These events, if confirmed, would represent a signal highly discrepant with astrophysical expectations, but also difficult to explain with dark matter within current experimental bounds. An annihilating dark matter model that produces detectable antihelium flux is in general expected to also produce antiprotons and antideuterons at large fluxes, the first of which has not been reported in great excess and the second of which has not been reported at all~\cite{Heeck:2019ego,Cirelli:2014qia,Blum:2017qnn,Korsmeier:2017xzj,Poulin:2018wzu}. However, these tentative events have motivated an interest in antihelium cosmic ray detection, and while projected sensitivities have not been reported by either experiment, it may still be informative to set theoretical expectations.

\begin{figure}
    \centering
    \begin{subfigure}[t]{0.5\textwidth}
        \centering
        \includegraphics[width=\linewidth]{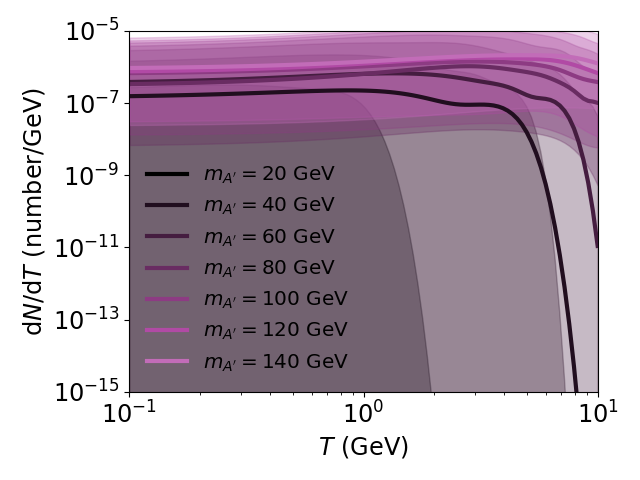}
    \end{subfigure}%
    ~ 
    \begin{subfigure}[t]{0.5\textwidth}
        \centering
        \includegraphics[width=\linewidth]{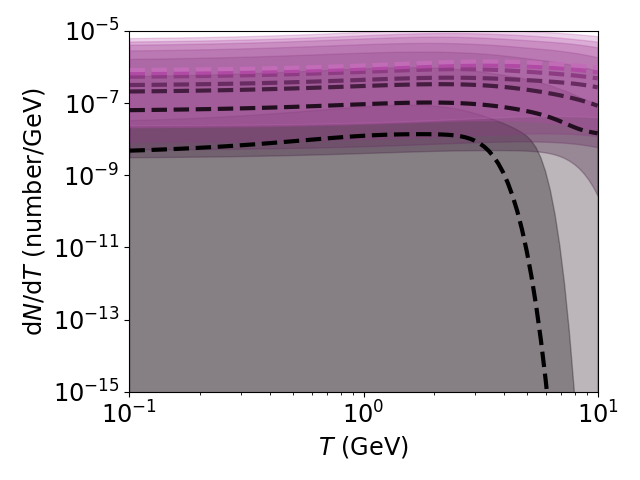}
    \end{subfigure}
    
        \begin{subfigure}[t]{0.5\textwidth}
        \centering
        \includegraphics[width=\linewidth]{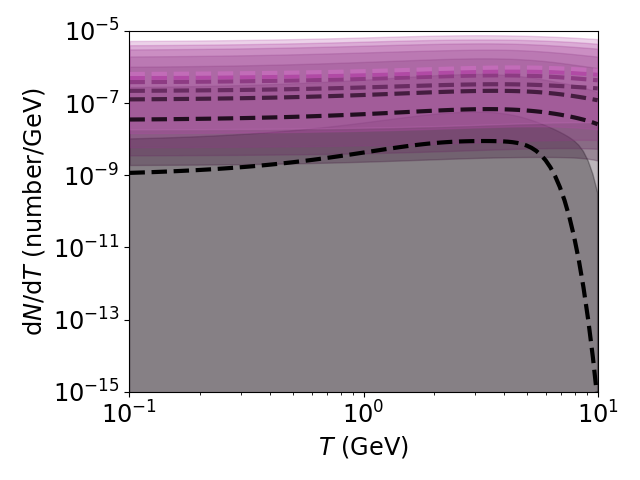}
    \end{subfigure}%
    ~ 
    \begin{subfigure}[t]{0.5\textwidth}
        \centering
        \includegraphics[width=\linewidth]{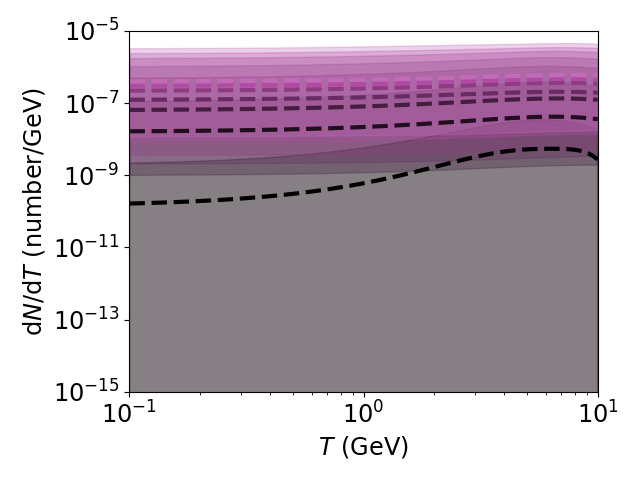}
    \end{subfigure}
    \caption{Injection spectra of antihelium for various dark photon masses, boosted to the rest frame of DM assuming $m_{A'}/m_\chi = 1$ (top left), $2/3$ (top right), $1/2$ (bottom left), and $1/3$ (bottom right). The line is evaluated at $p_0 = 225 \,\MeV$ and the shaded regions denote the range of injection spectra given by $125 \leq p_0 \leq 325\, \MeV$.}
    \label{fig:injspec_helium}
\end{figure}

The coalescence process of antihelium formation is understood even less well than that of antideuterons, as very little collider data has been reported on antihelium formation. Currently, only the ALICE experiment has published antihelium events, at $\sqrt{s} = 7$ TeV~\cite{Acharya:2017fvb}. Various coalescence models have been proposed~\cite{Cirelli:2014qia,Blum:2017qnn}, but here we adopt the afterburner formulation: an antinucleus is formed if the constituent antinucleons all lie within a sphere in momentum space of radius $p_0/2$ around the center-of-mass. Under this formulation fits to ALICE antihelium spectra using {\tt Pythia 8} have yielded $p_0\sim 225\, \MeV$. However, as we are considering center-of-mass energies much lower than these, we will study antihelium injection spectra for coalescence momenta $125\leq p_0 \leq 325 \, \MeV$. Due to computational cost, we simulate $10^8$ events for each center-of-mass energy ($m_{A'}$) we study, and report the cases with no antihelium events with an upper-bound. Figure~\ref{fig:injspec_helium} shows the injection spectra for various choices of $m_{A'}$ and $m_
\chi$. As expected, the antihelium injection is much weaker than antideuterons, by over three orders of magnitude.

\begin{figure}
    \centering
    \begin{subfigure}[t]{0.5\textwidth}
        \centering
        \includegraphics[width=\linewidth]{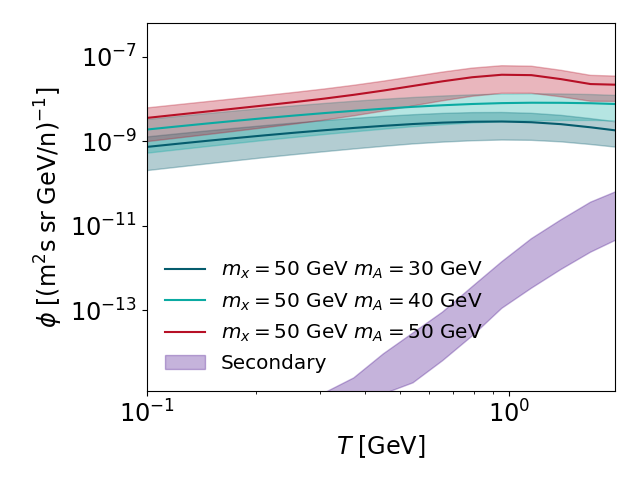}
    \end{subfigure}%
    ~ 
    \begin{subfigure}[t]{0.5\textwidth}
        \centering
        \includegraphics[width=\linewidth]{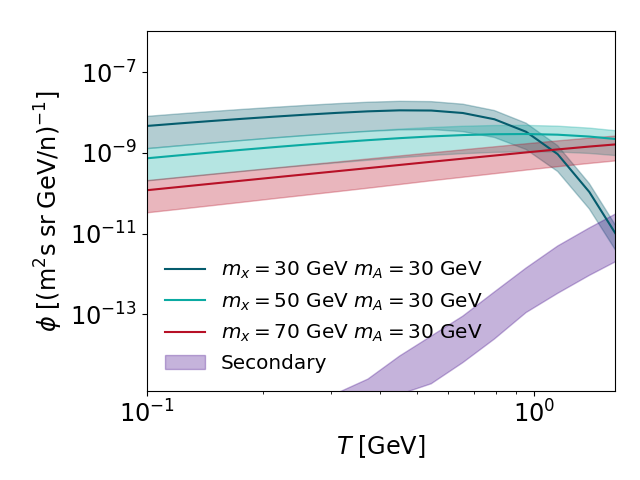}
    \end{subfigure}
    \caption{Predicted antihelium flux spectra of DPDM for various choices of $m_{A'}$ (left) and $m_\chi$ (right). The antihelium secondary is taken from~\cite{Korsmeier:2017xzj}, where the shaded region represents a range of coalescence momenta and the MED propagation parameters have been assumed.}
    \label{fig:flux_helium}
\end{figure}

\begin{figure}
    \centering
       \begin{subfigure}[t]{0.49\textwidth}
        \centering
        \includegraphics[width=\linewidth]{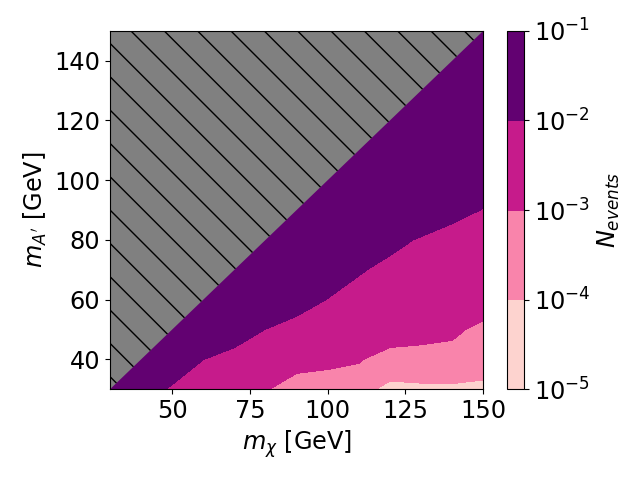}
    \end{subfigure}%
    ~ 
    \begin{subfigure}[t]{0.49\textwidth}
        \centering
        \includegraphics[width=\linewidth]{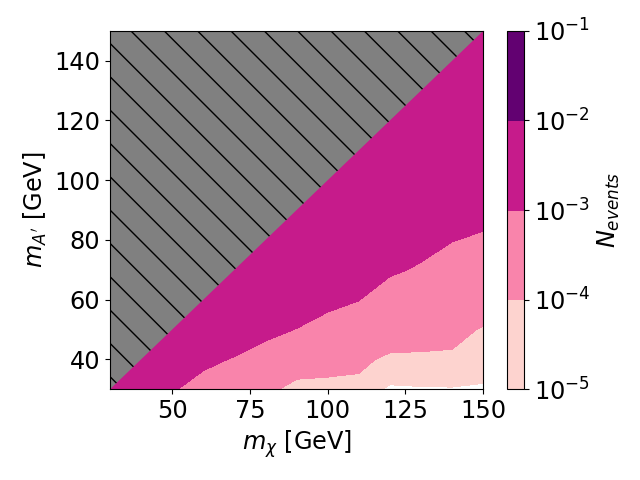}
    \end{subfigure}
    \caption{Expected number of antihelium events seen by \GAPS during the 105-day flight period, assuming the most optimistic MAX (left) and conservative MED (right) propagation model. As shown, the $O(10)$ events reported by \AMS is highly unrealizable with thermal relic cross sections.}
    \label{fig:Nevents_helium}
\end{figure}

The transport of antihelium cosmic rays through the ISM is modeled in a similar way to the antideuteron case, though cross-sections for antinuclei disintegration from collisions must be modified. We use the ISM propagation functions $R_{\bar{\mathrm{He}}}$ presented in Ref.~\cite{Cirelli:2014qia} to obtain the expected antihelium cosmic ray flux at the top of the atmosphere from DPDM annihilations. These are shown for various choices of $m_\chi$ and $m_{A'}$ in Fig.~\ref{fig:flux_helium}.  The expected antihelium secondary is shown for a range of coalescence momenta~\cite{Korsmeier:2017xzj}, and as shown the flux from DM annihilation is several orders of magnitude larger than the astrophysical background. However, the antihelium flux from DPDM is also shown to be comparable or lower than the expected antideuteron secondary in Fig.~\ref{fig:flux}. While the \GAPS detection sensitivities for antihelium are not well known, we expect them to be comparable to that for antideuteron, and we remind the reader that the expected secondary contribution there is  $\lesssim0.04$ events.  Fig.~\ref{fig:Nevents_helium} further shows the expected number of events for a range of masses assuming a thermal relic annihilation cross section in the conservative (right) and optimistic (left) propagation scenarios, and assuming antideuteron detection sensitivities.  If \GAPS sensitivities for antihelium detection are comparable to that of antideuteron detection, to obtain a $O(10)$ $\overline{\text{He}}$ event measurement with annihilating DPDM one would need a cross section two orders of magnitude above thermal relic for the most optimistic propagation model, which is difficult to realize particularly in absence of corresponding antideuteron events. Thus, we do not expect DPDM to be able to explain the tentative \AMS detections. However, if the antihelium detection occurs with larger efficiency than that of antideuterons, an $O(1)$ event measurement is conceivable within this model and would still be significantly in excess of the astrophysical expectation.

\section{Conclusions}
\label{sec:concl}

Thermal relic freeze-out remains the easiest way to explain the abundance of dark matter in the present day universe. However, increasingly stringent constraints from direct detection experiments exclude naive WIMP-type models and indicate that any accurate thermal relic model must annihilate via a hidden sector, such as dark photon dark matter, or otherwise have direct coupling with baryons greatly suppressed. Thus, the remaining viable way to search for these models is indirect detection. In contrast with indirect channels such as gamma rays and cosmic ray positrons or antiprotons, low energy cosmic ray antideuterons represent a unique probe of dark matter annihilations due to its very low astrophysical background. 

In this work we investigate the detection prospects of dark matter annihilating through a massive dark photon with cosmic ray antideuterons. The dominant source of uncertainty from this analysis derives from the choice of cosmic propagation and coalescence models, each of which induce an uncertainty of a factor of a few. The choice of halo profile in the propagation analysis represents a subdominant source of error. We show that for both optimistic and conservative propagation, for thermal relic dark matter $\langle \sigma |v| \rangle_{ann} = 1 \pb$, near-future experiment \GAPS is expected to measure signals from annihilating dark photon dark matter at the $3-5\sigma$ level for $m_\chi \sim 10-100\, \GeV$ for $m_\chi \approx m_{A'}$. The region of parameter space $m_chi \approx m_{A'} \sim 10 -40 \, \GeV$ favored by the \textit{Fermi} Galactic center excess measurement in particular is expected to be detected or ruled out at the $3\sigma$ level by \textit{GAPS}. 

\acknowledgments
We thank Prateek Agrawal, David Pinner, and James Unwin for illuminating discussions, and thank Rene Ong for insights into the \GAPS experiment. LR is supported by NSF grants PHY-1620806 and PHY-1915071, the Chau Foundation HS Chau postdoc support award, the Kavli Foundation grant ``Kavli Dream Team'', and the Moore Foundation Award 8342.

\bibliography{dm_indirect_detect,experiments_and_observations,dm_direct_detect,dm_collider,unsorted}
\bibliographystyle{JHEP}
\end{document}